\documentclass[11pt]{amsart}

\usepackage{amsmath}
\usepackage{amssymb}
\usepackage{graphicx}
\numberwithin{equation}{section}
\usepackage{color,stmaryrd,placeins}
\newcommand*{\pheq}{\mathrel{\phantom{=}}}
\newcommand*{\Eq}[1]{(\ref{eq:#1})}
\newcommand*{\Sec}[1]{\mbox{Sec.~\ref{sec:#1}}}
\newcommand*{\Fig}[1]{\mbox{Fig.~\ref{fig:#1}}}
\newcommand*{\Tbl}[1]{\mbox{Table~\ref{tbl:#1}}}
\newcommand*{\ds}{\displaystyle}
\newcommand*{\D}{\mathrm{d}}
\newcommand*{\Exp}[1]{\mathnormal{\,\mathrm{e}}^{\mbox{\footnotesize$#1$}}}
\newcommand*{\AND}{\quad\mbox{and}\quad}
\newcommand*{\WITH}{\quad\mbox{with}\quad}
\newcommand*{\FOR}{\quad\mbox{for}\quad}
\newcommand*{\OR}{\quad\mbox{or}\quad}
\newcommand*{\Max}[2][]{\mathop{\mathrm{Max}}_{#1}{\left\{#2\right\}}}
\newcommand*{\abs}[1]{\mathopen{\boldsymbol{|}}#1\mathclose{\boldsymbol{|}}}
\DeclareMathAlphabet{\vecfont}{OT1}{cmr}{bx}{it}
\renewcommand*{\vec}[1]{\vecfont{#1}}
\newcommand*{\grad}{\boldsymbol{\nabla}}
\title[Thomas--Fermi equation and Majorana]%
{Thomas--Fermi equation revisited:\\ %
  A variation on a theme by Majorana}

\author[B.-G. Englert]{Berthold-Georg Englert}

\address%
{School of Physics, Beijing Institute of Technology, %
  Beijing 100081, China {\normalfont and}
  Department of Physics, National University of Singapore, %
  Singapore 117551, Singapore
}
\email{\texttt{berge@bit.edu.cn, phyebg@nus.edu.sg}}

\subjclass[2010]{34A05, 35Q99, 53C29, 81V45 }
\keywords{Thomas--Fermi equation, Majorana's scaling transformation, homology}

\dedicatory{Dedicated to Goong Chen on the occasion of his 75th birthday.}

\begin{document}

\begin{abstract}
Majorana found a way to exploit the scaling properties of the Thomas--Fermi
equation for converting this second-order differential equation into one of
first order.
We explore his method for the familiar neutral-atom solution and extend it to
the solution that is relevant for weakly ionized atoms.
Various integrals and other quantities with importance for atomic physics are
recalculated and their values compared with the ones obtained in the 1980s by
more tedious numerical procedures.
\end{abstract}

\maketitle

\vspace*{-3ex}
\centerline{\small(Posted on the arXiv on 23 April 2026.)}
\vspace*{3ex}

\section{Introduction}\label{sec:intro}
In 1926 and 1927, respectively, Thomas \cite{Thomas:1927} and Fermi
\cite{Fermi:1928} managed to combine concepts of the old quantum theory
(before mid-1925, that is) with the then-recent methods of quantum mechanics
--- which grew rapidly from the seminal works of Hei\-sen\-berg,
Schr\"o\-dinger, and others --- for a first successful model of many-electron
atoms. 
The central mathematical object, the second-order differential equation now
known as the Thomas--Fermi (TF) equation, caught Majorana's attention, who had
the profound insight that there are scaling properties that can be exploited
for establishing an equivalent first-order differential equation. 

Most unfortunately, Majorana did not communicate his observations but,
fortunately, they were published more than sixty years after his mysterious
disappearance, namely when his private notes appeared in print thanks to the
commendable effort by Esposito, Majorana Jr., van der Merwe, and Recami
\cite{Esposito+3:2003}.
Moreover, Esposito provided a pedagogical exposition of Majorana's method,
supplemented by additional technical details \cite{Esposito:2002}.

In this contribution, we further elaborate on Majorana's first-order equation
and use it to compute various numbers with relevance for atomic
physics --- numbers that were obtained with more tedious methods before;
we refer, in particular, to the monograph of 1988 \cite{LNP300} and the works
cited therein.
In addition to the much studied solution of the TF equation of the original
model for neutral atoms, there is another solution that is important for weakly
ionized atoms;
we apply Majorana's procedure to that case, too, to find and then explore the
corresponding Majorana-type first-order equation.

We begin with a brief account of Thomas's and Fermi's reasoning in \Sec{T+F}.
Then, in \Sec{setstage}, we set the stage by recalling the second-order TF
equation and introducing three scale-invariant functions that are constructed
from the solutions of the TF equation.
Next, in \Sec{majo1}, we establish Majorana's first-order equation by an
argument that is closer in spirit to that in \cite{Alizzi+1:2024} than to the
original one in Majorana's notes, as recalled by
Esposito~\cite{Esposito:2002}.
The analogous argument for the second solution of the TF equation is presented
in \Sec{majo2}, where we report the other Majorana-type first-order equation.

Sections \ref{sec:num1} and \ref{sec:num2} deal with the power series for the
two equations, respectively.
The first series is the one that Esposito introduced \cite{Esposito:2002} and
the second is constructed in close analogy.
Both series are used for computing anew the basic numbers associated with the
TF functions, which are then compared with the numbers found in the 1980s. 
Further applications, in particular the computation of various relevant
integrals involving the neutral-atom TF function, are the subject matter of
\Sec{integrals}.
These are then used in \Sec{energy} for calculating anew the coefficients in
the formulas for the binding energy of neutral atoms and also the leading term
in the expression for the ionization energy.

We close with a summary and outlook in \Sec{sum-out}.

\section{How Thomas and Fermi reasoned}\label{sec:T+F}
Thomas and Fermi, independently, wondered about many-electron atoms, which
were beyond the reach of the then-new quantum mechanics.
The Schr\"odinger wave function for $N$ electrons has $2^N$ components, each a
com\-plex-valued function of the $3N$ electron coordinates.
A compact analytical expression for the atomic ground-state wave
function was out of the question then, as it is now, and there was no method
for obtaining meaningful approximations.%
\footnote{Hartree's first step in this direction \cite{Hartree:1928} %
  was still in the future.}

Thomas and Fermi resorted to modeling the atom by picturing the electrons
moving independently under the influence of the force associated with an
effective potential energy $V(\vec{r})$, a real-valued function of an
electron's position vector $\vec{r}$, which they approximated by
\begin{equation}\label{eq:TF0.a}
  V(\vec{r})=-\frac{Ze^2}{r}+\frac{Ze^2}{r_{\mathrm{ch}}^{\ }}
  +e^2\int(\D\vec{r}')\,\frac{n(\vec{r'})}{\abs{\vec{r}-\vec{r}'}}\,,
\end{equation}
where $e$ is the elementary charge, ${r=\abs{\vec{r}}}$ is the length of
$\vec{r}$, $(\D\vec{r}')$ denotes the euclidean volume element associated with
$\vec{r}'$, and $n(\vec{r}')$ is the density of electrons at $\vec{r}'$;
the first term is the Coulomb potential of the nucleus with charge
$Ze$, the second term is the chemical potential parameterized by the distance
$r_{\mathrm{ch}}^{\ }$, and the third term is the electrostatic potential of
the charge density $-en(\vec{r})$ of the electrons.
Note that, if the wave function were at hand, one would obtain $n(\vec{r})$ by
integrating the sum of the absolute squares of all components over the
coordinates of ${N-1}$ electrons.

The second ingredient is the statistical-physics expression for the electron
density in terms of the momentum integral over the region where the
single-electron energy $\vec{p}^2/(2m)+V(\vec{r})$ is negative,%
\footnote{As regulated by the chemical potential.}
\begin{align}\label{eq:TF0.b}
  n(\vec{r})
  &=\frac{2}{(2\pi\hbar)^3}\int(\D\vec{p})\;
  \Biggl\{
  \begin{array}{cl}
    1 & \ds\text{where}\ \vec{p}^2+2mV(\vec{r})<0\\[0.5ex]
    0 & \text{elsewhere}
  \end{array}\Biggr\}\\ \nonumber
  &=\frac{1}{3\pi^2\hbar^3}\Bigl(-2mV(\vec{r})\Bigr)_+^{3/2}\,,
\end{align}
counting two electrons per phase-space volume $(2\pi\hbar)^3$;
here, $m$ is the electron mass, $\vec{p}$ is the electron's momentum vector,
$2\pi\hbar$ is Planck's constant, and
${\bigl(y\bigr)_+=\bigl(y+\abs{y}\bigr)/2}$ suppresses negative values of the
real variable~$y$.
The approximations entering \Eq{TF0.a} and \Eq{TF0.b} are semi-classical in
nature; both can be refined by including quantum corrections, which have been
studied extensively in the century since Thomas's and Fermi's seminal work.

Upon applying the Laplace differential operator on both sides of \Eq{TF0.a},
\begin{equation}\label{eq:TF0.c}
  r>0\,:\quad-\grad^2V(\vec{r})=4\pi e^2 n(\vec{r})\,,
\end{equation}
and taking into account that an isolated atom is isotropic by writing
\begin{equation}\label{eq:TF0.d}
  V(\vec{r})=-\frac{Ze^2}{r}f(\kappa r)
  \quad\text{with}\quad
  \kappa\frac{\hbar^2}{me^2}=8(6\pi)^{-2/3}Z^{1/3}\,,
\end{equation}
we arrive at the familiar TF differential equation for $f(x)$ in \Eq{TF1}
below. 
The combination of physical constants that multiplies $\kappa$ is Bohr's radius,
the natural unit of length in atomic physics, and the numerical factor is
approximately $61/54$.

The first steps taken by Thomas and Fermi triggered a large body of work in
the next two decades, summarized in Gomb\'as's classic of
1949~\cite{Gombas:1949}.
In the systematic development of density functional theory from the mid-1960s
onward, the TF model is the basic approximation, and one has learned, in which
sense the TF approximation is exact when both $N$ and $Z$ are very large.
Recent accounts on the state of affairs from the physical and the
mathematical perspectives are the chapters by Okun and Burke
\cite{OkunBurke:2023} and Siedentop \cite{Siedentop:2023}, respectively, in
the proceedings of a workshop in~2019.

\section{Setting the stage}\label{sec:setstage}
In the TF equation, 
\begin{equation}\label{eq:TF1}
  f''(x)={\left\{
  \begin{array}{c@{\text{\ \ when\ \ }}l}
    \ds x^{-1/2}f(x)^{3/2}& f(x)\geq0\,,\\[1ex]
                      0 & f(x)\leq0\,,
  \end{array}\right.}
\end{equation}
the variable $x$ is positive with ${x=0}$ included for some solutions and the
function values $f(x)$ are real.
For later reference, we note that
\begin{align}\label{eq:TF-id}
  7x^{-1/2}f(x)^{5/2}
  &=\frac{\D}{\D x}\Bigl(5f(x)f'(x)
       -5xf'(x)^2+4xf(x)f''(x)\Biggr)\,,\\ \nonumber
  7f'(x)^2
  &=\frac{\D}{\D x}\Bigl(2f(x)f'(x)
       +5xf'(x)^2-4xf(x)f''(x)\Bigr)\,,\\ \nonumber
  12f(x)^4-13f'(x)^3
  &=\frac{\D}{\D x}\Bigl(12xf(x)f'(x)f''(x)-3f(x)f'(x)^2-10xf'(x)^3\Bigr)
\end{align}
are valid identities for all solutions of \Eq{TF1} where ${f(x)\geq0}$.

\begin{figure}[!t]
  \centering
  \includegraphics[viewport=130 540 385 715,clip]{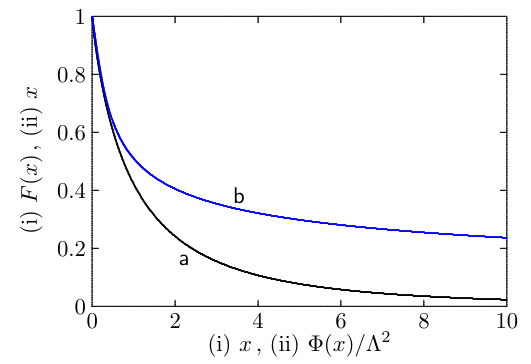}
  \caption{\label{fig:1} The TF functions of the two kinds.
  Curve \textsf{a} is the graph of
  ${x\mapsto F(x)}$ for \mbox{case~(i)} and curve \textsf{b} is the graph of
  ${\Phi(x)/\Lambda^2\mapsfrom x}$ for \mbox{case~(ii)}.}
\end{figure}

We are interested in two particular solutions of the TF equation,
which are specified by
\begin{align}\label{eq:TF2a}
  (\textrm{i})\  0\leq x<\infty:
  \quad f(0)&=1\,,\quad f'(0)=-B\,,\\\nonumber
  \phantom{ x\geq0\,:\quad}
  f(x)&=\frac{144}{x^3}\Bigl(1-\beta
     x^{-\gamma}+\cdots\Bigr)\mbox{\ for\ } x\gg1\,,%\nonumber\\[1ex]
\intertext{and}\label{eq:TF2b}  
  (\textrm{ii})\  0< x\leq1:
  \quad   f(1)&=0\,,\quad f'(1)=-\Lambda^2\,,
        \\ \nonumber
         \phantom{ 0< x\leq1\,:\quad}f(x)&=\frac{144}{x^3}\Bigl(1-\alpha
    x^{\sigma}+\cdots\Bigr)\mbox{\ for\ }0<x\ll1\,,
\end{align}
where ${\gamma=\bigl(\sqrt{73}-7\bigr)/2}$ and
${\sigma=\bigl(\sqrt{73}+7\bigr)/2}$ and the ellipses stand for higher powers 
of $x^{-\gamma}$ or $x^{\sigma}$, respectively.
We know from previous work (see \cite{LNP300} and the references therein)
that\footnote{In statements like these, it is understood that the initial
  digits of the decimal representations of real numbers are given.
  For the physics applications in \Sec{energy}, six digits are enough.}
\begin{equation}\label{eq:TF3}
  \begin{array}[t]{rrclrcl}
  (\textrm{i})\ & B&=&1.588\,071\,022\,61\,,&
                  \beta&=&13.270\,973\,848\,,\\
  (\textrm{ii})\ & \Lambda&=&32.729\,416\,116\,173\,,&\quad
                   \alpha&=&1.040\,180\,657\,3862
  \end{array}
\end{equation}
are valid digits.
We write $F(x)$ for the solution of case (i), which is relevant for the TF
model of neutral atoms, and $\Phi(x)$ for the solution of case (ii), which
informs us about weakly ionized atoms.
See \Fig{1} for the graphs of the two particular solutions of the TF equation.

%%%%
%%%%

If $x\mapsto f(x)$ solves the TF equation \Eq{TF1}, then
$x\mapsto
f_{\lambda}(x)=\lambda^3f(\lambda x)$ with $\lambda>0$ is another solution.
Following Majorana's guidance \cite{Esposito:2002}, we consider three
scale-invariant combinations,%
\footnote{Or homology-invariant combinations \cite{Alizzi+1:2024}, %
  if you prefer this terminology.} namely
\begin{align}\label{eq:M1a}
    P(x)&=x^{3/2}f(x)^{1/2}\,,
     &
 \D P&=\frac{\D x}{x}
       \biggl(\frac{3}{2}P-\frac{1}{2}P^{-1}Q\biggr)\\\nonumber
     &&&=\frac{\D x}{x}
       \biggl(\frac{3}{2}P-\frac{1}{2}P^{5/3}R\biggr)\,,
        \\\label{eq:M1b}
 Q(x)&=-x^4f'(x)\,,
     &
       \D Q&=\frac{\D x}{x}\Bigl(4Q-P^3\Bigr)\\\nonumber
     &&&
       =\frac{\D x}{x}\biggl(4Q-Q^{9/8}R^{-9/8}\biggr)\,,
             \\\label{eq:M1c}
 R(x)&=-f(x)^{-4/3}f'(x)\,,
     &
 \D R&=\frac{\D x}{x}\biggl(\frac{4}{3}R^2P^{2/3}-P^{1/3}\Biggr)\\\nonumber
     &&&=\frac{\D x}{x}\biggl(\frac{4}{3}R^{7/4}Q^{1/4}
         -R^{-1/8}Q^{1/8}\Biggr)
\end{align}
with $Q^3=R^3P^8$.
They are scale-invariant in the sense that the replacement $f\to f_{\lambda}$
yields $P(x)\to P(\lambda x)$,  $Q(x)\to Q(\lambda x)$, and
$R(x)\to R(\lambda x)$ without additional powers of $\lambda$ as overall
factors.
In \Eq{M1a} and \Eq{M1c}, ${f(x)>0}$ is assumed; for ${f(x)\leq0}$,
put ${P(x)=0}$ and ${R(x)^{-1}=0}$.

We wish to find one of $P,Q,R$ as a function of another and
choose the pair in accordance with the following endpoint values:
\begin{equation}\label{eq:M2}
  \begin{array}[t]{@{}rl|ccc}
    &&P&Q&R\\ \hline\rule{0pt}{12pt}%
    (\textrm{i})&x\to0&0&0&B\\
    &x\to\infty\;&\ 12\ &\ 432\ &\ (3/16)^{1/3}\\[2ex]
    (\textrm{ii})&x\to0&12&432& (3/16)^{1/3}\\
    &x\to1&0&\Lambda^2&\infty
  \end{array}
\end{equation}
so that we want $R$ as a function of $P$ in case (i) and $Q$ as a function of
$P$ in case (ii).

\section{Majorana's equation for case ({\normalfont i})}\label{sec:majo1}
\begin{figure}[!t]
  \centering
  \includegraphics[viewport=130 550 365 715,clip]{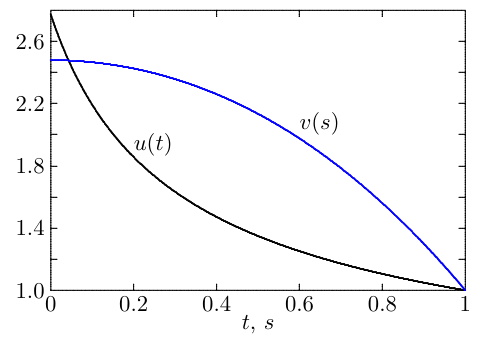}
  \caption{\label{fig:2}Graphs of the functions ${t\mapsto u(t)}$ and
    ${s\mapsto v(s)}$. }
\end{figure}

In case (i), we look for $R$ as a function of $P$.
As a consequence of \Eq{M1a} and \Eq{M1c}, we have the differential equation
\begin{equation}\label{eq:M3}
   \frac{\D R}{\D P}
   =-\frac{2}{3}P^{-2/3}\,\frac{3-4P^{1/3}R^2}{3-P^{2/3}R} \,.
\end{equation}
Keeping the values in \Eq{M2} in mind, we put
\begin{equation}\label{eq:M4}
  P=12t^3\,,\quad R%=3(12)^{-2/3}u(t)
  =\biggl(\frac{3}{16}\biggr)^{1/3}u(t)
\end{equation}
and arrive at Majorana's equation \cite[eq.~(27)]{Esposito:2002},
\begin{equation}\label{eq:M5a}
  \frac{\D u(t)}{\D t}=-8\frac{1-tu(t)^2}{1-t^2u(t)}\,,
\end{equation}
with
\begin{equation}\label{eq:M5b}
  u(1)=1\,,\quad u(0)=\biggl(\frac{16}{3}\biggr)^{1/3}B\quad\mbox{and}\quad
  u'(1)=-2(1-\gamma)\,;%=-\bigl(9-\sqrt{73}\,\bigr)\,.
\end{equation}
see \Fig{2} for the graph of ${t\mapsto u(t)}$.

%%%%
%%%%

We then find $x$, $F(x)$, and $F'(x)$ in terms of $t$ and $u(t)$ from
\begin{align}\label{eq:M5c}
  \frac{\D x}{x}
  &=\D P\,\biggl(\frac{3}{2}P-\frac{1}{2}P^{5/3}R\biggr)^{-1}\\\nonumber
  &=2\frac{\D t}{t}\,\frac{1}{1-t^2u(t)}
  =2\frac{\D t}{t}+2\,\D t\,\frac{tu(t)}{1-t^2u(t)}
\end{align}
so that \cite[eqs.~(42) and (43)]{Esposito:2002}
\begin{equation}\label{eq:M5d1}
  x=\lambda t^2\Exp{2U(t)}\,,\quad
  F(x)=\frac{144}{\lambda^3}\Exp{-6U(t)}\,,\quad
  -F'(x)=\frac{432}{\lambda^4}u(t)\,\Exp{-8U(t)}
\end{equation}
with ${\lambda^3=144}$ and
\begin{equation}\label{eq:M5d2}
   U(t)=\int_0^t\D t'\,\frac{t'u(t')}{1-{t'}^2u(t')}\,,
\end{equation}
where the value of the scale parameter $\lambda$ is determined by
the requirement of $F(x=0)=1$ in \Eq{TF2a}.

Together, then, we obtain the case-(i) solution of the second-order
TF differential equation in \Eq{TF1} by solving Majorana's first-order
differential equation in \Eq{M5a} and evaluating the integral in \Eq{M5d2}.
While the reduction from second to first order is an obvious simplification
in itself, there is the additional great benefit that the Majorana function
${t\mapsto u(t)}$ has a power series that converges in the whole range of
${0\leq t\leq1}$ (see \cite{Esposito:2002} and \Sec{num1}),
whereas the two power series for the TF function ${x\mapsto F(x)}$ have
finite, non-overlapping intervals of convergence \cite{LNP300}.%
\footnote{More specifically, the series in powers of $x^{-\gamma}$ converges
  for ${x\gtrsim30}$ and that in powers of $x^{1/2}$ converges for
  ${x\lesssim0.16}$.} 

\begin{figure}[!t]
  \centering
  \includegraphics[viewport=130 545 365 715,clip]{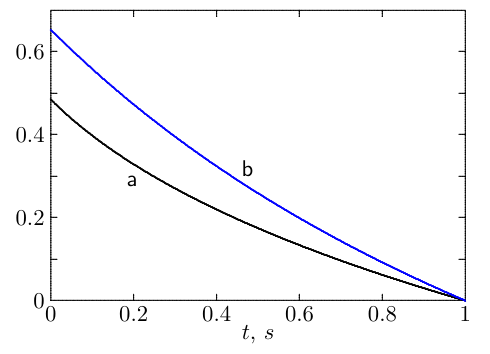}
  \caption{\label{fig:3} The sums in \Eq{SA2} and \Eq{SB2a}
    as functions of $t$ or~$s$.
  Curve \textsf{a} is the graph of
  ${t\mapsto2\gamma\bigl(U(t)+\overline{U}\bigr)+\log(1-t)}$;
  curve \textsf{b} is the graph of
  ${s\mapsto(2\sigma/3)\bigl(V(s)+\overline{V}\bigr)+\log(1-s)}$.}
\end{figure}

Now, turning to $x\gg1$, we have
\begin{align}\label{eq:M5e}
  \beta=\biggl(1-\frac{x^3F(x)}{144}\biggr)x^{\gamma}\Biggr|_{x\to\infty}
        &=\Bigl(1-t^6\Bigr)\Bigl(\lambda t^2\Exp{2U(t)}\Bigr)^{\gamma}
          \Biggr|_{1>t\to1}\\ \nonumber
         &=6(12)^{2\gamma/3}\,\Exp{-2\gamma \overline{U}}
\end{align}
with
\begin{equation}\label{eq:M5f}
  \overline{U}
  =\int_0^1\D t\,\biggl(\frac{1}{2\gamma}\frac{1}{1-t}
  -\frac{tu(t)}{1-t^2u(t)}\biggr)\,.  
\end{equation}
See \Fig{3} for the graph of
${t\mapsto2\gamma\bigl(U(t)+\overline{U}\bigr)+\log(1-t)}$.%
\footnote{Here and in later occurrences, $\log(\ )$ denotes the natural
  logarithm.} 

%%%%
%%%%

\section{The analog of Majorana's equation for case ({\normalfont ii})}
\label{sec:majo2}%
In case (ii), we have the differential equation
\begin{equation}\label{eq:M6}
  \frac{\D Q}{\D P}
    =-2\frac{4PQ-P^4}{Q-3P^2}
\end{equation}
for finding $Q$ as a function of $P$.
We put
\begin{equation}\label{eq:M7}
  P=12s\,,\quad
  Q=432v(s)
\end{equation}
and obtain
\begin{equation}\label{eq:M8a}
  \frac{\D v(s)}{\D s}=-\frac{8}{3}\frac{sv(s)-s^4}{v(s)-s^2}
\end{equation}
with
\begin{equation}\label{eq:M8b}
  v(1)=1\,,\quad v(0)=\frac{1}{432}\Lambda^2\quad\mbox{and}\quad
  v'(1)=-\frac{2}{3}(\sigma-3)\,;%=-\frac{1}{3}\bigl(\sqrt{73}+1\bigr)\,.
\end{equation}
see \Fig{2} for the graph of $s\mapsto v(s)$.

We then find $x$, $\Phi(x)$, and $\Phi'(x)$ in terms of $s$ and $v(s)$ from
\begin{equation}\label{eq:M8c}
  \frac{\D x}{x}
  =\D P\,\biggl(\frac{3}{2}P-\frac{1}{2}P^{-1}Q\biggr)^{-1}
  =-\frac{2}{3} \frac{\D s\,s}{v(s)-s^2}
\end{equation}
so that \enlargethispage{1.3\baselineskip}  %%++
\begin{align}\label{eq:M8d1}
  &x=\Exp{-(2/3)V(s)}\,,\hspace*{-4em}
  &\Phi(x)&=144s^2\,\Exp{2V(s)}\,,\\\nonumber
   && -\Phi'(x)&=432v(s)\,\Exp{(8/3)V(s)}
\end{align}
with
\begin{equation}\label{eq:M8d2}
   V(s)=\int_0^s\D s'\,\frac{s'}{v(s')-{s'}^2}\,,
\end{equation}
where no additional scale factor is needed to ensure that $x\to1$ when
$s\to0$.

The remarks about case~(i) in the paragraph between \Eq{M5d2} and \Eq{M5e} are
equally valid for case~(ii):
The second-order TF equation is converted into the first-order equation in
\Eq{M8a} and the integration in \Eq{M8d2}, whereby ${s\to v(s)}$ has a power
series that converges for all $s$ in the range ${0\leq s\leq1}$ (see
\Sec{num2}), whereas the 
two power series for ${x\to \Phi(x)}$ have finite, yet overlapping, intervals
of convergence \cite{LNP300}.%
\footnote{More specifically, the series in powers of $x^{\sigma}$ converges
  for ${x\lesssim0.94}$ and that in powers of $(1-x)^{1/2}$ converges for
  ${x\gtrsim0.88}$.}   

Now, turning to $0<x\ll1$, we have
\begin{align}\label{eq:M8e}
  \alpha=\biggl(1-\frac{x^3\Phi(x)}{144}\biggr)x^{-\sigma}\Biggr|_{0<x\to0}
          &=\Bigl(1-s^2\Bigr)\Exp{(2\sigma/3)V(s)}\Biggr|_{1>s\to1} \\ \nonumber
          &=2\Exp{-(2\sigma/3)\overline{V}}
\end{align}
with
\begin{equation}\label{eq:M8f}
  \overline{V}=
  \int_0^1\D s\,\biggl(\frac{3}{2\sigma}\frac{1}{1-s}
                       -\frac{s}{v(s)-s^2}\biggr)\,.
\end{equation}
See \Fig{3} for the graph of 
${s\mapsto(2\sigma/3)\bigl(V(s)+\overline{V}\bigr)+\log(1-s)}$.

\section{Numerical procedure for case ({\normalfont i})}\label{sec:num1}
\begin{figure}
  \centering
  \includegraphics[viewport=134 545 434 715,clip]{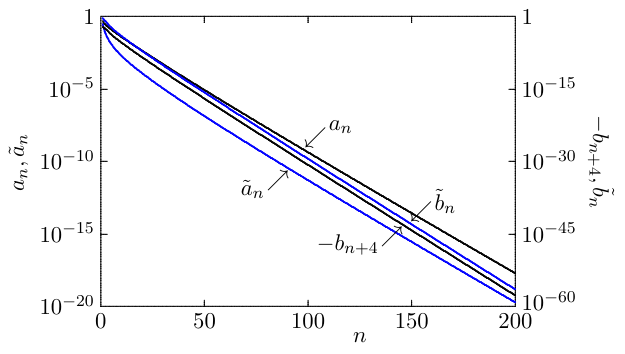}
  \caption{\label{fig:4}Lin-log plot of the coefficients in \Eq{SA1},
    \Eq{SA2}, \Eq{SB1}, and \Eq{SB2a} for ${1\leq n\leq200}$
    (or $5\leq n\leq204$ for $b_n$).
  The \mbox{large-$n$} slopes are those of geometric sequences with
  convergence for ${\abs{1-t}\lesssim1.20}$ and ${\abs{1-s}\lesssim1.84}$,
  respectively, for the corresponding power series. }
\end{figure}

Esposito \cite{Esposito:2002} notes that an expansion of $u(t)$ in powers
of $1-t$ is useful, 
\begin{align}\label{eq:SA1}
       u(t)&=\sum_{n=0}^{\infty}a_n(1-t)^n  \\\nonumber
  \WITH a_0&=1 \AND a_1=-u'(1)=2-2\gamma=9-\sqrt{73}\,,
\end{align}
and reports the recurrence relation for the $a_n$s, which is equivalent to
\begin{align}\label{eq:SA1''}      
  a_n&=\Bigl(2n+16-(n+1)a_1\Bigr)^{-1}\\ \nonumber
     &\pheq{}\times\Biggl(
       \frac{1}{2}(n+15)\sum_{m=0}^{n-1}a_ma_{n-1-m}
       -(n+8)\sum_{m=1}^{n-1}a_ma_{n-m}\\ \nonumber
  &\pheq{}\hphantom{\times\Biggl(}\ 
       +\frac{1}{2}(n+1)\sum_{m=2}^{n-1}a_ma_{n-1-m}\Biggr)
\end{align}
for ${n>1}$, where the third sum is empty for ${n=2}$, when
\begin{equation}\label{eq:SA1'}
  a_2=\frac{17-10a_1}{20-3a_1}a_1=\frac{6497-755\sqrt{73}}{152}\,.
\end{equation}
For their values, see \cite[Table~1]{Esposito:2002}
or the entries in \Tbl{1} and the lin-log graph in \Fig{4}; the
graph of $u(t)$ in \Fig{2} results from this expansion. 

%%%%
\begin{table}
  \centering
  \caption{\label{tbl:1}The first $20$ coefficients in the power series
    for $u(t)$ in \Eq{SA1}, $U(t)$ in \Eq{SA2},
    $v(s)$ in \Eq{SB1}, and $V(s)$ in \Eq{SB2a}.}
  \begin{tabular}{rrrrr}
   \hline\hline\rule{0pt}{14pt}%
    $n$&\multicolumn{1}{c}{$a_n$}&\multicolumn{1}{c}{$\tilde{a}_n$} %
    &\multicolumn{1}{c}{$b_n$}&\multicolumn{1}{c}{$\tilde{b}_n$}
    \\ \hline\rule{0pt}{12pt}%
  1 & 0.4559\,9625 & 0.2898\,1876 &   5.1813\,3458 & 0.4290\,2657 \\
  2 & 0.3044\,5508 & 0.0901\,7378 &$-$2.9584\,0435 & 0.1483\,3053 \\
  3 & 0.2221\,7976 & 0.0409\,2810 &   0.2678\,6614 & 0.0492\,0658 \\
  4 & 0.1682\,1262 & 0.0221\,2613 &   0.0021\,4562 & 0.0168\,5127 \\
  5 & 0.1298\,0407 & 0.0132\,4507 &$-$0.0070\,8457 & 0.0061\,1999 \\[0.5ex]
  6 & 0.1013\,0020 & 0.0084\,6769 &$-$0.0037\,2358 & 0.0023\,9137 \\
  7 & 0.0796\,3516 & 0.0056\,6326 &$-$0.0015\,1117 & 0.0010\,0172 \\
  8 & 0.0629\,2304 & 0.0039\,1218 &$-$0.0005\,7357 & 0.0004\,4236 \\
  9 & 0.0499\,0531 & 0.0027\,6832 &$-$0.0002\,2091 & 0.0002\,0213 \\
 10 & 0.0396\,9618 & 0.0019\,9536 &$-$0.0000\,9025 & 0.0000\,9430 \\[0.5ex]
 11 & 0.0316\,4979 & 0.0014\,5926 &$-$0.0000\,3946 & 0.0000\,4458 \\
 12 & 0.0252\,8385 & 0.0010\,7976 &$-$0.0000\,1817 & 0.0000\,2129 \\
 13 & 0.0202\,3221 & 0.0008\,0668 &$-$0.0000\,0862 & 0.0000\,1026 \\
 14 & 0.0162\,1359 & 0.0006\,0754 &$-$0.0000\,0415 & 0.0000\,0498 \\
 15 & 0.0130\,1011 & 0.0004\,6071 &$-$0.0000\,0201 & 0.0000\,0244 \\[0.5ex]
 16 & 0.0104\,5183 & 0.0003\,5144 &$-$0.0000\,0098 & 0.0000\,0120 \\
 17 & 0.0084\,0559 & 0.0002\,6946 &$-$0.0000\,0048 & 0.0000\,0060 \\
 18 & 0.0067\,6661 & 0.0002\,0755 &$-$0.0000\,0024 & 0.0000\,0030 \\
 19 & 0.0054\,5216 & 0.0001\,6051 &$-$0.0000\,0012 & 0.0000\,0015 \\
 20 & 0.0043\,9678 & 0.0001\,2458 &$-$0.0000\,0006 & 0.0000\,0007 \\
 \hline\hline\rule[-14pt]{0pt}{10pt}
  \end{tabular}
\end{table}
%%%%

%%%%
%%%%

There is a related power series for the integrand in \Eq{M5d2} and \Eq{M5f},
\begin{align}\label{eq:SA2}
  \frac{\D}{\D t}U(t)
  &=\frac{tu(t)}{1-t^2u(t)}=\frac{1}{2\gamma}\frac{1}{1-t}
  {\left(1-\sum_{n=1}^{\infty}n\tilde{a}_n(1-t)^n\right)}\,,\\\nonumber
  U(t)&=\frac{1}{2\gamma}{\left(-\log(1-t)
                                +\sum_{n=1}^{\infty}\tilde{a}_n(1-t)^n
                                -\sum_{n=1}^{\infty}\tilde{a}_n\right)}\,,
\end{align}
with
\begin{equation}
    n\tilde{a}_n=a_n'-a_n''+\sum_{m=1}^{n-1}m\tilde{a}_ma''_{n-m}\,, 
\end{equation}
where
\begin{align}
  a_1'=1-a_1\,,\quad
  a_1''=\bigl(a_2-2a_1\bigr)/(2\gamma)
\end{align}
and
\begin{align}
  a'_k=a_{k-1}-a_k\,,\quad
  a_k''=\bigl(a_{k+1}-2a_k+a_{k-1}\bigr)/(2\gamma)\FOR k>1\,;
\end{align}
see \Tbl{1} and \Fig{4}.
The latter informs us that both series converge for all $t$
in the whole range of $0\leq t\leq1$ so that \cite[eq.~(39)]{Esposito:2002}
\begin{equation}\label{eq:SA3a}
  \biggl(\frac{16}{3}\biggr)^{1/3}B=u(0)=\sum_{n=0}^{\infty}a_n
\end{equation}
and
\begin{equation}\label{eq:SA3b}
  \log\bigl(6(12)^{2\gamma/3}/\beta\bigr)
  =2\gamma\overline{U}=\sum_{n=1}^{\infty}\tilde{a}_n\,.
\end{equation}
All the digits in \Eq{TF3} are confirmed in this way by the algorithm
described in this section, which is much easier to implement than the one that
yielded the values in the 1980s.

Obviously, each coefficient $a_n$ is the sum of a rational number and a
rational multiple of $\sqrt{73}$, and so is every coefficient $\tilde{a}_n$.
Therefore, one can obtain exact expressions for the $a_n$s and the
$\tilde{a}_n$s by writing the rational ingredients as ratios of integers, as
exemplified by $a_0$, $a_1$, $a_2$ above.
Indeed, Moses computed the first 5,000 digits of $B$ in this way
\cite{Moses:2013}.
We are less ambitious and content with the decimal approximations obtained
with double precision arithmetic, for which ${1+2^{-52}>1+2^{-53}=1}$, so that
single-step rounding-off errors affect the 16th digit,
inasmuch as $2^{-53}=1.11\times10^{-16}$.

\section{Numerical procedure for case ({\normalfont ii})}\label{sec:num2}
The analogous power series for $v(s)$ is
\begin{align}\label{eq:SB1}
  v(s)&=s^2+\sum_{n=1}^{\infty}b_n(1-s)^n  \\\nonumber
  \WITH b_1&=2-v'(1)=\frac{2}{3}\sigma=\frac{1}{3}\bigl(\sqrt{73}+7\bigr)
\end{align}
and\footnote{Here, $\delta_{a,b}$ is the Kronecker delta symbol.}
\begin{align}\label{eq:SB2}
  b_n&=-\biggl((n+1)b_1-\frac{14}{3}\biggr)^{-1} \\\nonumber
             &\mathrel{\phantom{=-}}\times{\left(\frac{14}{3}b_{n-1}+
      8\delta_{n,2}-8\delta_{n,3}+\frac{8}{3}\delta_{n,4}
      +\frac{1}{2}(n+1)\sum_{m=2}^{n-1}b_mb_{n+1-m}\right)}
\end{align}
for ${n>1}$ with an empty sum for ${n=2}$, when
\begin{equation}\label{eq:SB2'}
  b_2=-\frac{14b_1+24}{9b_1-14}=-\frac{469+103\sqrt{73}}{456}\,.
\end{equation}
See \Tbl{1} and \Fig{4};
the graph of $v(s)$ in \Fig{2} results from this expansion.
The corresponding series for the integrand in \Eq{M8d2} and \Eq{M8f} is
\begin{align}\label{eq:SB2a}
  \frac{\D}{\D s}V(s)
  &=\frac{s}{v(s)-s^2}=\frac{3}{2\sigma}\frac{1}{1-s}
  {\left(1-\sum_{n=1}^{\infty}n\tilde{b}_n(1-s)^n\right)}\,,\\\nonumber
  V(s)&=\frac{3}{2\sigma}{\left(-\log(1-s)
      +\sum_{n=1}^{\infty}\tilde{b}_n(1-s)^n
      -\sum_{n=1}^{\infty}\tilde{b}_n\right)}
\end{align}
with
\begin{align}\label{eq:SB2b}
  \tilde{b}_1&=1+b_2/b_1\\\nonumber \AND
  n\tilde{b}_n&=b_{n+1}/b_1-\sum_{m=1}^{n-1}m\tilde{b}_mb_{n+1-m}/b_1
  \FOR n>1\,;
\end{align}
see \Tbl{1} and \Fig{4}.
Here, too, we have convergence for the whole range of ${0\leq s\leq1}$  so that
\begin{equation}\label{eq:SB3a}
  \frac{1}{432}\Lambda^2=v(0)=\sum_{n=1}^{\infty}b_n
\end{equation}
and
\begin{equation}\label{eq:SB3b}
  \log(2/\alpha)=\frac{2\sigma}{3}\overline{V}
  =\sum_{n=1}^{\infty}\tilde{b}_n\,,
\end{equation}
with all the digits in \Eq{TF3} confirmed.

\section{Integrals involving the TF function and other matters}
\label{sec:integrals}%
\subsection{Case (i)}
The locus and value of the maximum of $xF(x)$,
\begin{align}\label{eq:OA1}
  &\Max[x]{xF(x)}=x_0F(x_0)=0.486\,348\,5380 \\ \nonumber
  \WITH  &x_0=2.104\,025\,2802 \AND F(x_0)=0.231\,151\,4708 \,,
\end{align}
follow after finding $t_0=0.496\,342\,166\,063$ as the solution of
${3t^2u(t)=1}$.
A simple search algorithm is repeated bisection of an interval that contains
$t_0$, perhaps starting with ${0<t_0<1}$.

Integrals of products of powers of $x$ and powers of $F(x)$ can be evaluated in
accordance with
\begin{equation}\label{eq:OA2a}
  \int_0^{\infty}\D x\,x^kF(x)^l
  =\frac{6l\,(12)^{2(k+1)/3}}{3l-k-1}
    \int_0^1\D t\,t^{2k+1}\Exp{-2(3l-k-1)U(t)}
\end{equation}
for ${3l>k+1>0}$.
The conversion of the $x$ integral to the $t$ integral for the
parameterization of \Eq{M5d1} and \Eq{M5d2} employs
\begin{align}\label{eq:OA2b}
  &\pheq\D\biggl(\frac{x^{k-3l+1}}{k-3l+1}\bigl(x^3F(x)\bigr)^l\biggr)
    \\ \nonumber
  &=\D x\,x^kF(x)^l
    -\frac{x^{-(3l-k-1)}}{3l-k-1}\D\Bigl(\bigl(x^3F(x)\bigr)^l\Bigr)
    \\ \nonumber
  &=\D x\,x^kF(x)^l
    -\D t\,\frac{x^{-(3l-k-1)}}{3l-k-1}\frac{6l}{t}\bigl(x^3F(x)\bigr)^l
    \\ \nonumber
  &=\D x\,x^kF(x)^l
    -\D t\,\frac{6l\,(12)^{2(k+1)/3}}{3l-k-1}\,
    t^{2k+1}\Exp{-2(3l-k-1)U(t)}\,,
\end{align}
where ${x^3F(x)=144 t^6}$ is taken into account; the boundary terms of the
integration by parts do not contribute.

For benchmarking, we compute the $t$ interal in
\begin{align}\label{eq:OA3}
  u(0)=\biggl(\frac{16}{3}\biggr)^{1/3}B
  &=\frac{7}{5}\biggl(\frac{16}{3}\biggr)^{1/3}
    \int_0^{\infty}\D x\,x^{-1/2}F(x)^{5/2} \\ \nonumber
  &=12\int_0^1\D t\;\Exp{-14U(t)}\,,
\end{align}
which follows from the first or the second identity in \Eq{TF-id}.
We find
\begin{align}
  u(0)&=2.774\,615\,643\,934\,192 \quad \mbox{from this $t$ integral,}\\
  \nonumber
  u(0)&=2.774\,615\,643\,934\,217 \quad \mbox{from \Eq{SA3a},}
\end{align}
and infer that such $t$ integrals yield thirteen correct digits.
Other tests, which exploit
\begin{align}\label{eq:OA3a}
  1=\bigl(xF'(x)-F(x)\bigr)\Bigr|_{x=0}^{\infty}
  &=\int_0^{\infty}\D x\,x^{1/2}F(x)^{3/2}\\ \nonumber
  &=36\int_0^1\D t\,t^2\,\Exp{-6U(t)}
\intertext{or}
\label{eq:OA3b}
     u(0)=\biggl(\frac{16}{3}\biggr)^{1/3}B
  &=\biggl(\frac{16}{3}\biggr)^{1/3}\int_0^{\infty}\D x\,x^{-1/2}F(x)^{3/2}
  \\ \nonumber
  &=9\int_0^1\D t\;\Exp{-8U(t)}\,,
\end{align}
confirm this observation.

The exponential functions of multiples of $U(t)$ in \Eq{OA2a} and \Eq{OA3} to
\Eq{OA3b} are handled with the aid of  
\begin{align}\label{eq:OA4}
  \Exp{-2\kappa U(t)}&=\Exp{2\kappa\overline{U}}(1-t)^{\kappa/\gamma}
   {\left(1-\sum_{n=1}^{\infty}\overline{a}_n(\kappa/\gamma)(1-t)^n\right)}
   \\\nonumber
  \phantom{\Exp{-2\kappa U(t)}}&=(6/\beta)^{\kappa/\gamma}(12)^{2\kappa/3}
    (1-t)^{\kappa/\gamma}
   {\left(1-\sum_{n=1}^{\infty}\overline{a}_n(\kappa/\gamma)(1-t)^n\right)}
   \\\nonumber
  \WITH \overline{a}_1(\mu)&=\mu\tilde{a}_1\\\nonumber
  \AND  \overline{a}_n(\mu)&=\mu\tilde{a}_n
        -\frac{1}{n}\sum_{m=1}^{n-1}\overline{a}_m(\mu)(n-m)\mu\tilde{a}_m
        \FOR n>1\,.
\end{align}
For the $\kappa$ values we need most,
the first ten $\overline{a}_n(\kappa/\gamma)$
values are given in \Tbl{2}.
\Tbl{3} lists values of the integrals of $t^{2k+1}\Exp{-2(3l-k-1)U(t)}$, 
and the resulting values of the various integrals of products of powers of $x$
and $F(x)$ are reported in \Tbl{4};
the last row refers to the integral of $\bigl(-F'(x)\bigr)^3$, which is related
to that of $F(x)^4$ by the third identity in \Eq{TF-id},
so that
\begin{equation}\label{eq:OA6}
  \int_0^{\infty}\D x\,\Bigl(-F'(x)\Bigr)^3
  =\frac{3}{13}B^2-\frac{12}{13}\int_0^{\infty}\D x\,F(x)^4\,.
\end{equation}

%%%%
\begin{table}
  \centering\tabcolsep=4pt
  \caption{\label{tbl:2}%
    The first ten coefficients $\overline{a}_n(\kappa/\gamma)$ for
    ${\kappa=1,2,5,7,\text{\ or\ }11}$.}\vspace*{-10pt}
  \begin{tabular}{rrrrrr}\hline\hline\rule{0pt}{12pt}%
    $n$& \multicolumn{1}{c}{$\kappa=1$}
    & \multicolumn{1}{c}{$\kappa=2$}
    & \multicolumn{1}{c}{$\kappa=5$}
    & \multicolumn{1}{c}{$\kappa=7$}
    & \multicolumn{1}{c}{$\kappa=11$}\\[2pt]
    \hline\rule{0pt}{12pt}%
    1 & 0.3754\,1199 &   0.7508\,2398 &   1.8770\,5995 &   2.6278\,8392
      &   4.1295\,3188 \\
    2 & 0.0463\,3805 &$-$0.0482\,5807 &$-$1.1776\,5138 &$-$2.6352\,5107
      &$-$7.2416\,6037 \\
    3 & 0.0179\,8357 &   0.0011\,7541 &   0.2710\,8440 &   1.2470\,5199
      &   7.0141\,5650 \\
    4 & 0.0093\,3964 &   0.0030\,2958 &$-$0.0131\,9263 &$-$0.2727\,3723
      &$-$4.0799\,4849 \\
    5 & 0.0055\,3367 &   0.0023\,8827 &$-$0.0004\,8152 &   0.0205\,9164
      &   1.4458\,5545 \\[0.5ex]
    6 & 0.0035\,2913 &   0.0017\,1448 &$-$0.0000\,4477 &   0.0001\,6081
      &$-$0.2997\,1995 \\
    7 & 0.0023\,5782 &   0.0012\,1712 &$-$0.0000\,0506 &   0.0000\,1105
      &   0.0319\,4957 \\
    8 & 0.0016\,2628 &   0.0008\,6893 &   0.0000\,0077 &$-$0.0000\,0111
      &$-$0.0011\,4625 \\
    9 &  0.0011\,4805 &   0.0006\,2623 &   0.0000\,0211 &$-$0.0000\,0192
      &$-$0.0000\,1360 \\
   10 & 0.0008\,2490 &   0.0004\,5576 &   0.0000\,0247 &$-$0.0000\,0152
      &$-$0.0000\,0071 \\
    \hline\hline\rule[-14pt]{0pt}{10pt}
  \end{tabular}\tabcolsep=6pt
\end{table}
%%%%

\subsection{Case (ii)}
A particular integral of $\Phi(x)$ is relevant, namely
\begin{align}\label{eq:OB1}
  &\pheq-\frac{1}{4(12)^4}\int_0^1\D x\,\frac{1}{x^5}\,
  \frac{\D}{\D x}\Bigl(x^6\Phi(x)^2\Bigr)\\\nonumber
    &=\int_0^1\D s\,\frac{s^3}{x^5}
  =\int_0^1\D s\,s^3\,\Exp{(10/3)V(s)}=1.056\,061\,2411\,,
\end{align}
which is computed in analogy with \Eq{OA4} from the ingredients in \Eq{SB2a},
\Eq{SB2b}, and \Eq{SB3b}.
The 1980s value is $\ds(5/4)\times0.844\,849=1.056\,061$; all digits are
confirmed. 

%%%%
\begin{table}[!t]
  \centering
  \caption{\label{tbl:3} Values of
    ${\ds\int_0^1\D t\,t^{2k+1}\Exp{-2(3l-k-1)U(t)}}$ for ${3l>k+1>0}$.}%
  \vspace*{-10pt}
  \begin{tabular}{crrrr}\hline\hline\rule{0pt}{12pt}%
    $l$ & \multicolumn{1}{c}{$k=-1/2$}
        & \multicolumn{1}{c}{$k=0$}
        & \multicolumn{1}{c}{$k=1/2$}
        & \multicolumn{1}{c}{$k=1$}\\[2pt]
    \hline\rule{0pt}{12pt}%
  1/2 & 0.57014\,68753 & 0.28246\,66572 &                &                \\
  1   & 0.38752\,34122 & 0.11447\,54905 & 0.06381\,79642 & 0.05577\,72153 \\
  3/2 & 0.30829\,06271 & 0.06927\,85149 & 0.02777\,77778 & 0.01614\,68088 \\
  2   & 0.26212\,56210 & 0.04892\,33930 & 0.01587\,35703 & 0.00723\,68823 \\%
    [0.5ex]
  5/2 & 0.23121\,79703 & 0.03750\,88969 & 0.01041\,66667 & 0.00399\,58174 \\
  3   & 0.20876\,26928 & 0.03026\,40417 & 0.00743\,39529 & 0.00249\,48104 \\
  7/2 & 0.19154\,29334 & 0.02528\,25781 & 0.00561\,26138 & 0.00168\,92551 \\
  4   & 0.17782\,11892 & 0.02165\,98536 & 0.00441\,21461 & 0.00121\,15438 \\%
    [2pt]\hline\hline\rule[-14pt]{0pt}{10pt}
  \end{tabular}
\end{table}
%%%%

%%%%
\begin{table}[!t]
  \caption{\label{tbl:4} Integrals of the TF function for case (i).
    The first column shows  ${\kappa=3l-k-1}$.
    The third column lists the values computed in the 1980s with the correct
    digits underlined. The slanted number 3 in the last row marks a typo in
    \cite{LNP300} that was not noticed before.}\vspace*{-10pt}
  \begin{tabular}{ccll}\hline\hline\rule{0pt}{12pt}%
    &integrand in & 1980s value & now computed \\
    $\kappa$ & $\int_0^\infty\D x\,(\cdots)$ & in Ref.~\cite{LNP300}
    & with \Eq{OA2a}, \Eq{OA4} \\[2pt]
    \hline\rule{0pt}{12pt}%
 2& $F(x)$                 & \underline{1.800\,063\,94} & 1.800\,063\,9396\\
 5& $F(x)^2$               & \underline{0.615\,434\,6}4 & 0.615\,434\,6934\\
11& $F(x)^4$               &                            & 0.247\,701\,2723\\%
    [0.5ex]    
 1& $xF(x)$                & \underline{9.194}          & 9.194\,252\,0822\\
 1& $\bigl(F(x)/x\bigr)^{1/2}$  & \underline{3.915\,93}3 & 3.915\,931\,4911\\
  & $\bigl(-F'(x)\bigr)^3$ & \underline{0.353}\,\textsl{3}\,\underline{345\,6}
                           & 0.353\,345\,6501\\[2pt]
   \hline\hline\rule[-14pt]{0pt}{10pt}
  \end{tabular}
\end{table}
%%%%

\section{Energy formulas}\label{sec:energy}
In atomic units ($1\,\mathrm{Ha}=27.211\,386\,246\,\mathrm{eV}$),
the binding energy of a neutral atom with ${N=Z}$ electrons is
\begin{equation}\label{eq:E1}
  -E=c_7Z^{7/3}-\frac{1}{2}Z^2
      +c_5Z^{5/3}+\cdots
\end{equation}
with the coefficients 
\begin{align}\label{eq:E2}
  c_7&=\frac{6}{7}(3\pi/4)^{-2/3}B
  =\frac{6}{7}(3\pi^2)^{-1/3}u(0)=0.768\,745\,124\,,
    \\ \nonumber
  c_5&=\frac{11}{8}(3\pi/4)^{-4/3}
    \int_0^{\infty}\D x\,F(x)^2 \\ \nonumber
     &=\frac{264}{5}\bigl(3\pi^2\bigr)^{-2/3}
       \int_0^1\D t\;t\,\Exp{-10U(t)}
       =0.269\,900\,170\,.
\end{align}
The ellipsis in \Eq{E1} stands for other contributions to the binding energy.
First, there is an oscillatory term that is dominated by
\begin{equation}\label{eq:E3}
  -E_{\textrm{osc}}=c_4\bigl\lceil{c_{\textrm{p}}Z^{1/3}}\bigr\rfloor
  \biggl(\frac{1}{4}-\bigl\lceil{c_{\textrm{p}}Z^{1/3}}\bigr\rfloor^2\biggr)
  Z^{4/3}+\cdots\,, 
\end{equation}
where $\lceil y\rfloor$ denotes the difference between $y$ and its nearest
integer, and 
\begin{align}\label{eq:E4}
  c_4&=4(3\pi/4)^{-2/3}\frac{F(x_0)}{x_0}
       \biggl(1-\frac{1}{2}x_0\sqrt{x_0F(x_0)}\biggr)^{-1/2}=0.320\,593\,992\,,
   \\ \nonumber
  c_{\textrm{p}}&=(3\pi/4)^{1/3}\sqrt{x_0F(x_0)}=0.927\,991\,901
\end{align}
for the amplitude and the period.
Second, there is a relativistic correction with the leading terms
\begin{equation}\label{eq:E5}
  -E_{\textrm{rel}}=Z^2(Z\alpha^{\ }_{\textrm{fs}})^2
  \Bigl(c^{\textrm{rel}}_0-c^{\textrm{rel}}_1Z^{-1/3}
  +c^{\textrm{rel}}_2Z^{-2/3}+\cdots\Bigr)\,,
\end{equation}
where $\alpha^{\ }_{\textrm{fs}}=1/137.035\,999\,084$ is Sommerfeld's fine
structure constant and the coefficients are
\begin{align}\label{eq:E6a}
  c^{\textrm{rel}}_0
  &=\frac{5\pi^2}{24}-\zeta(3)=0.854\,110\,680\,,
      \\ \nonumber
  c^{\textrm{rel}}_1
  &=3(3\pi/4)^{-4/3}{\left(B^2
     -\int_0^{\infty}\D x\,\bigl(-F'(x)\bigr)^3\right)}
   \\ \nonumber
  &=\frac{\bigl(3\pi^2\bigr)^{-2/3}}{13}{\left(30u(0)^2
    +\frac{(24)^3}{11}\int_0^1\D t\,t\,\Exp{-22U(t)}\right)}
   \\ \nonumber
  &=2.075\,028\,109\,,
   \\ \nonumber
  c^{\textrm{rel}}_2
  &=2(3\pi/4)^{-2/3}B
    =2\bigl(3\pi^2\bigr)^{-1/3}u(0)
    = 1.793\,738\,623\,.
\end{align}
When relativistic contributions are ignored, the first ionization energy for
$Z\gg1$ is 
\begin{align}\label{eq:E7}
  I&=\frac{6}{7}\biggl(\frac{3\pi}{4}\Lambda\biggr)^{-2/3}
     +\frac{11}{10}\biggl(\frac{3\pi}{4}\Lambda\biggr)^{-4/3}
     \biggl(\frac{144}{\Lambda}\biggr)^2
     \int_0^1\D s\,s^3\,\Exp{(10/3)V(s)}\\[1ex] \nonumber
   &=0.047\,310\,0726+0.068\,506\,3874 \\ \nonumber
  &=0.115\,816\,460 \mathrel{\widehat{=}}3.151\,526\, \mathrm{eV}\,,
\end{align}
where the integral in \Eq{OB1} enters.

\section{Summary and outlook}\label{sec:sum-out}
We supplemented Majorana's first-order differential equation for the
neu\-tral-atom TF function with the analogous equation for the TF function for
weakly ionized atoms, and explored both equations with the aid of Esposito's
power series and its analog.
This provided a rather simple way of computing various numbers of physical
significance, thereby confirming the values found in the 1980s by much more
tedious numerical calculations.

The TF equation is a particular case of
\begin{align}\label{eq:so1}
  &f''(x)=x^af(x)^b \OR f''(x)=-x^af(x)^b \\ \nonumber
  \WITH &f_{\lambda}(x)=\lambda^{(a+2)/(b-1)}f(\lambda x)\,.
\end{align}
Other nonlinear (${b\neq1}$) equations of this form can be investigated as
well, among them the Lane--Emden equation for polytropic gas balls (negative
sign and ${a+b=1}$); see \cite{Alizzi+1:2024} in this context and the
references therein.

In the context of the TF equation itself, it remains to be seen how Majorana's
transformation helps in studying the solutions specified by
\begin{equation}\label{eq:so2}
  f(0)=1\,,\quad f(x_{\mathrm{ch}}^{\ })=0\,,\quad
  -x_{\mathrm{ch}}^{\ }f'(x_{\mathrm{ch}}^{\ })=q \FOR 0<q<1\,,
\end{equation}
where the value of the positive $x_{\mathrm{ch}}^{\ }$ depends on ${q=1-N/Z}$,
the degree of ionization.%
\footnote{The relation ${x_{\mathrm{ch}}^{\ }=q\kappa r_{\mathrm{ch}}^{\ }}$
  connects with \Eq{TF0.d}.
  The two cases in \Eq{TF2a} and \Eq{TF2b}
  are obtained in the limit of ${q\to0}$, ${x_{\mathrm{ch}}^{\ }\to\infty}$
  without and with the scaling for ${\lambda=x_{\mathrm{ch}}^{\ }}$,
  respectively.} 
This is unexplored territory.

\section*{Dedication}
It gives me great pleasure to dedicate these notes to Goong Chen with whom I
share a passion in algorithms that exploit scaling
properties~\cite{Chen+2:2004}.  

\section*{Acknowledgment}
I am extremely grateful for the long-standing support from the Centre for
Quantum Technologies, Singapore, where part of this work was done.

\end{document}